# Spin Splitting in Altermagnetic RuO$_2$ Enables Field-free Spin-Orbit Torque Switching via Dominant Out-of-Plane Spin Polarization

Zhuoyi Li[1,2,3,#], Zhe Zhang[1,2,3,#], Xianyang Lu[1,2,3,*], Yongbing Xu[1,2,3,4,*]

[1]*National Key Laboratory of Spintronics, Nanjing University, Suzhou 215163, China*

[2]*Jiangsu Provincial Key Laboratory of Advanced Photonic and Electronic Materials, School of Electronic Science and Engineering, Nanjing University, Nanjing 210093, China*

[3]*School of Integrated Circuits, Nanjing University, Suzhou 215163, China*

[4]*York-Nanjing International Center for Spintronics (YNICS), School of Physics, Engineering and Technology, University of York, York YO10 5DD, UK*

[*]Authors to whom correspondence should be addressed: xylu@nju.edu.cn and ybxu@nju.edu.cn
[#]Z. Li and Z. Zhang contributed equally to this work

# Abstract

Researchers have recently identified a novel class of magnetism, termed "altermagnetism", which exhibits characteristics of both ferromagnetism and antiferromagnetism. Here, we report a groundbreaking discovery of efficient field-free spin-orbit torque (SOT) switching in a $RuO_2$ (101)/Co/Pt/Co/Pt/Ta structure. Our results demonstrate that the spin current flows along the [100] axis, induced by the in-plane charge current, with the spin polarization direction aligned parallel to the Néel vector. These z-polarized spins generate an out-of-plane anti-damping torque, enabling deterministic switching of the Co/Pt layer without the necessity of an external magnetic field. The altermagnetic spin splitting effect (ASSE) in $RuO_2$ promotes the generation of spin currents with pronounced anisotropic behavior, maximized when the charge current flows along the [010] direction. This unique capability yields the highest field-free switching ratio, maintaining stable SOT switching within an external field range of approximately 400 Oe. Notably, ASSE dominates the spin current, especially when the current is aligned with the [010] direction (θ = 90°). Here, the spin polarization component $\sigma_z$ creates a substantial field-like effective field, surpassing the damping-like field from $\sigma_y$. This highlights the crucial role of $\sigma_z$ in enhancing spin-torque efficiency and elucidating spin flow modulation mechanics in this crystalline context. Our study highlights the potential of $RuO_2$ as a powerful spin current generator, paving the way for practical applications in spin-torque switching technologies and other cutting-edge spintronic devices.

Researchers have recently identified a novel class of magnetism that exhibits characteristics of both ferromagnetism and antiferromagnetism. This new phenomenon, predicted to occur in over 200 materials, has been termed "altermagnetism"[1,2]. Materials such as ruthenium dioxide ($RuO_2$) could exhibit this dual nature, combining the stable, fast spin-flipping properties of antiferromagnets with the distinct spin states of ferromagnets. Unlike typical materials where electron spins align with the atomic orientation in the crystal lattice, in altermagnets, spin arrows can rotate independently of the atoms. Previously considered a paramagnet, $RuO_2$ has been shown to exhibit itinerant antiferromagnetism, with a Néel temperature above 300 K and the Néel vector aligned along the [001] axis[3]. Recent theoretical work has suggested that collinear antiferromagnet $RuO_2$ might generate strong electric-field-induced spin currents with spin orientation roughly aligned along the Néel vector[4]. $RuO_2$ crystallizes in the rutile structure with the $P4_2/mnm$ space group, where ruthenium atoms are situated in the centers of stretched oxygen octahedrons[5]. This octahedral crystal field results in an anisotropic electronic structure and elliptical Fermi surfaces at $k_z = 0$[4]. The 90° rotation of Ru atoms in opposite magnetic sublattices, surrounded by directionally distinct oxygen octahedrons, leads to anisotropic spin band splitting in momentum space, making $RuO_2$ an efficient spin splitter[6].

H.Bai et al. have experimentally demonstrated spin splitting torque (SST) in collinear antiferromagnet $RuO_2$ films, showing that the spin current direction is correlated with the crystal orientation of $RuO_2$ and that the spin polarization direction is parallel to the Néel vector[6]. Furthermore, when the Néel vector is slightly canted, a strong out-of-plane spin current can result. Complementing this, Arnab Bose et al. have shown that (101)-oriented $RuO_2$ films can generate a significant electric-field-induced out-of-plane damping-like torque on an adjacent permalloy film[7]. Yaqin Guo et al. Have further shown the spin currents along the z-direction arising from the altermagnetic spin splitting effect (ASSE) show an anisotropic behavior and are maximized when an electric current is along the [010] direction[8]. This finding underscores the potential of $RuO_2$ to act as a powerful spin current generator, leveraging the ASSE in a collinear antiferromagnet. The spin current direction in these films is directly influenced by the crystal orientation, while the spin polarization direction is determined by the Néel vector of $RuO_2$. This controllability can significantly enhance the efficiency of out-of-plane spin polarization generation, offering a promising alternative to other mechanisms that rely on low crystal symmetry and specific magnetic ordering[9-11]. This characteristic is particularly important for the development of spin-orbit torque (SOT) devices, which are critical components in modern spintronic applications.

SOT devices have garnered considerable attention in the last decade for their potential in memory and logic applications, notably in spin-orbit torque magnetic random-access memory (SOT-MRAM)[12-14]. The ability to precisely control and efficiently generate spin currents is crucial for these devices, as it directly impacts their performance, energy efficiency, and scalability[15]. In this article, we report a method for achieving field-free SOT switching in perpendicular magnetic film structure grown on the (101)-oriented $RuO_2$ film. The z-polarized spins induce an out-of-plane anti-damping torque, enabling deterministic switching of the Co/Pt layer without the necessity of an external magnetic field. And it has been observed to have a clear dependence on the direction of $J_C$. The sample demonstrates an optimal switching ratio nearly 100% at an applied in-plane field $H_x$ = 0 Oe, when $J_C$ flowing along the [0$\bar{1}$0] axis. Notably, the ASSE dominates the spin current, especially when the applied current is aligned with the [010] crystallographic direction (θ = 90°). In this configuration, the spin polarization component $\sigma_z$ exerts a significant influence, creating a substantial field-like effective field that surpasses the damping-like effective field associated with $\sigma_y$. This interplay highlights the crucial role of $\sigma_z$ in enhancing spin-torque efficiency, thereby elucidating the mechanics of spin flow modulation within this crystalline context. The significance of these findings offering practical pathways to enhance the functionality and efficiency of spin torque switching technologies. This new avenue not only broadens the horizons of spintronic applications but also brings us closer to realizing advanced memory and logic devices that are both powerful and energy-efficient[16-18].

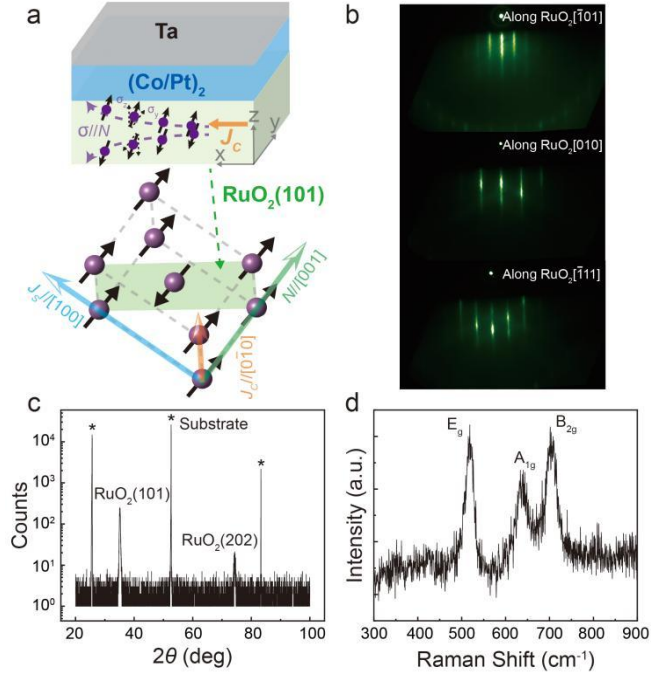

**Fig. 1. Charge-to-spin conversion via ASSE in RuO$_2$ (101). a,** Schematic representation of our sample. The spin polarization direction ($\sigma$) for the spin current ($J_S$) is aligned parallel to the Néel vector ([001] axis). The schematic illustrates $J_S$ flowing along the [100] axis induced by the charge current ($J_C$) along the [0$\bar{1}$0] axis. The RuO$_2$ (101) crystal plane is highlighted by the green shading. **b,** Reflection high-energy electron diffraction (RHEED) patterns of a 15 nm RuO$_2$ (101) film grown on the Al$_2$O$_3$ (1$\bar{1}$02) substrate along two different directions: [$\bar{1}$01] and [$\bar{1}$11]. **c,** $\theta$-$2\theta$ scan x-ray diffraction (XRD) spectrum of a 50 nm thick RuO$_2$ (101) film grown on the Al$_2$O$_3$(1$\bar{1}$02) substrate. Peaks from the substrate are marked with *. **d,** Raman spectra of a 50 nm thick RuO$_2$ (101) film grown on the Al$_2$O$_3$(1$\bar{1}$02) substrate, showing three Raman active modes: $E_g$, $A_{1g}$ and $B_{2g}$ modes.

In this work, we propose a RuO$_2$ (101)/Co/Pt/Co/Pt/Ta structure, which successfully achieves efficient field-free SOT switching. All samples were grown on Al$_2$O$_3$ (1$\bar{1}$02) substrates using a magnetron sputtering system. The layer sequence RuO$_2$ (15 nm)/Co (0.5 nm)/Pt (1 nm)/Co (0.5 nm)/Pt (1 nm)/Ta (2 nm) was deposited from bottom to top, as shown in Fig. 1a. The (101)-oriented RuO$_2$ generates spin current with out-of-plane spin polarization[19]. The spin current ($J_S$) flowing along the [100] axis, induced by the charge current ($J_C$) along the [0$\bar{1}$0] axis. The spin polarization direction ($\sigma$) for $J_S$ is approximately aligned parallel to the Néel vector

([001] axis)[6]. High-quality $RuO_2$ (101) films were grown on single-crystal $Al_2O_3$ ($1\bar{1}02$) substrates by introducing $O_2$ gas into an Ar base gas during film growth in our magnetron sputtering system. During the film growth process, a 50 standard cubic centimeter per minute (sccm) Ar gas flow was introduced, while the $O_2$ flow rate was controlled at 10 sccm. A pure ruthenium target was used, with the Radio Frequency (RF) power set at 50 W. The substrate temperature was fixed at 500 °C during deposition. As shown in Fig. 1b, the clear and sharp RHEED pattern for the 15 nm $RuO_2$ (101) film indicates that the film exhibits good crystallinity with a flat, well-ordered surface[20,21]. XRD measurements of a 50 nm $RuO_2$ film, shown in Fig. 1c, reveal clear and sharp $RuO_2$ (101) and $RuO_2$ (202) peaks, confirming the good (101) orientation of our $RuO_2$ films. Raman spectroscopy analysis, shown in Fig. 1d, indicates three Raman active modes for the $RuO_2$ (101) film: $E_g$, $A_{1g}$ and $B_{2g}$ modes[22,23]. Simultaneously, we tested the resistivity of the $RuO_2$ film. The I-V curve is provided in the supplementary materials, and the resistivity is 1.24 μΩ·m, which is consistent with the ultralow resistivity reported in single crystals in the literature, indicating excellent conductivity[24]. These results demonstrate the high quality and excellent structural properties of the $RuO_2$ (101) films grown in this study.

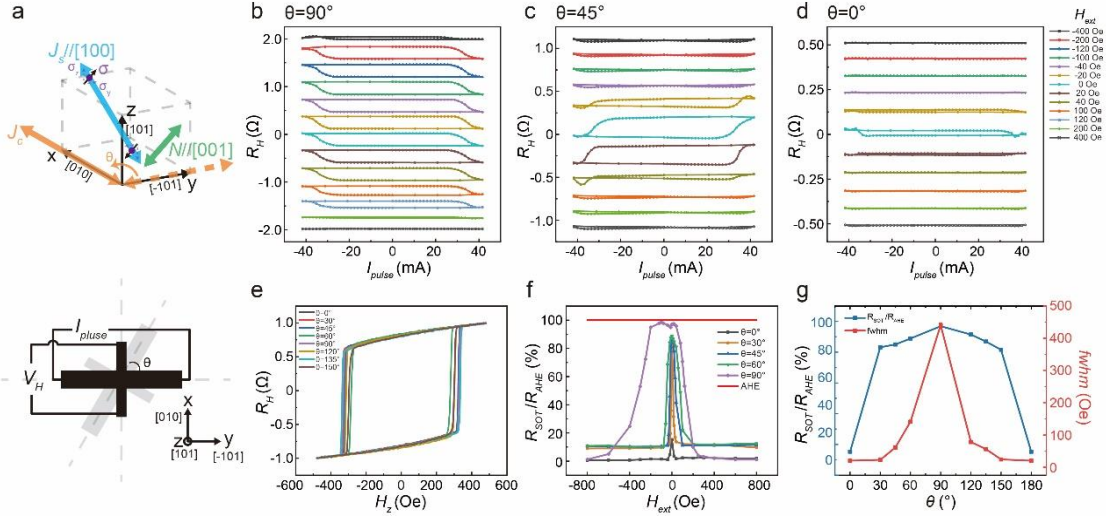

**Fig. 2. Current-driven SOT switching measurement in RuO$_2$ (101)/Co/Pt/Co/Pt. a,** Schematic diagram illustrating the generation of spin current via ASSE in the (101)-oriented RuO$_2$ film, along with the Hall device used for the experiments. θ represents the angle between the current direction and the y-axis. **b, c, d,** Corresponding current-induced magnetization switching behaviors observed in Co/Pt multilayers at θ = 90°, 45° and 0°. **e,** Anomalous Hall effect (AHE) loops for diffenent values of θ. **f,** The SOT switching ratios ($R_{SOT}/R_{AHE}$) under various θ and magnitudes of external magnetic field ($H_{ext}$). **g,** The field-free SOT switching ratio and the full width at half maximum (FWHM) of the switching ratio under different values of θ.

We initially investigate the current induced field-free SOT switching behavior. Fig. 2a shows schematic diagrams of the generation of spin current via ASSE in the (101)-oriented RuO$_2$ film, along with the Hall device used for the experiments. To achieve SOT switching, we swept a pulsed direct current and measured the Hall resistance change of the Hall bar (see Methods). Fig. 2b-d display the current-induced SOT magnetization switching loops in RuO$_2$/Co/Pt/Co/Pt at different in-plane external fields ($H_{ext}$) from + 400 Oe to - 400 Oe at θ values of 90°, 45°, and 0°, respectively, with $H_{ext}$ aligned parallel to the direction of the applied current ($I_{pulse}$). In conventional structures, no switching loop occurs at zero magnetic field, necessitating $H_{ext}$ to break the rotational symmetry of the spin torque. Remarkably, at θ = 90°, where the current flows along the x-axis ([010] crystal direction), SOT switching can be achieved without $H_{ext}$, with a switching ratio approaching 100%, as shown in Fig. 2b. Similarly, field-free SOT switching is observed at θ = 45° (Fig. 2c), whereas no magnetization switching occurs at θ = 0° (Fig. 2d). The SOT switching curves for other angles, such as 30°, 60°, 120°, 135°, and 150°, are presented in Supplementary Information. Fig. 2e compares the anomalous Hall effect (AHE) measurement curves at different θ, all of which exhibit good perpendicular anisotropy with coercive fields around 300 Oe. Fig. 2f summarizes the SOT switching ratios ($R_{SOT}/R_{AHE}$) under different θ and various magnitudes of $H_{ext}$, indicating that the highest SOT switching ratios are achieved without any external field assistance, regardless of the current direction. The maximum ratio is near 100% at θ = 90° and the minimum at θ = 0°. Additionally, the switching direction remains unchanged regardless of the magnitude of $H_{ext}$. This behavior is significantly different from traditional SOT magnetization switching curves[25-27]. Fig. 2g summarizes the field-free SOT switching ratio and the full width at half maximum (FWHM) of the switching ratio under different values of θ. It is evident that θ = 90° achieves the highest field-free switching ratio and FWHM, maintaining stable SOT switching within an external field range of approximately 400 Oe, indicating excellent resistance to magnetic interference under these conditions. The corresponding values for other angles are symmetrical about θ = 90° within the range of 0° to 180°, gradually decreasing away from θ = 90°. This indicates that the spin currents along the [100] direction with an out-of-plane spin polarization component $\sigma_z$, arising from the ASSE, exhibit anisotropic behavior and are maximized when the charge current flows along the [010] direction. These findings are consistent with previous reports in the literature[8,19].

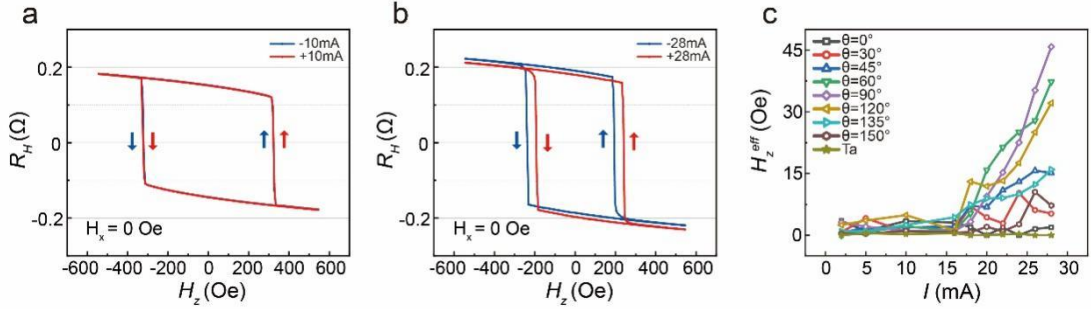

**Fig. 3 Current-induced effective fields with z-polarized spin in RuO$_2$ (101). a-b,** The AHE loops with the applied current of ± 10 mA and ± 28 mA on RuO$_2$/Co/Pt/Co/Pt, respectively at an in-plane field, $H_x$ = 0 Oe. **c,** A summary of the shift ($\Delta H_z^{eff}$) at different bias currents (*I*) for RuO$_2$/Co/Pt/Co/Pt without $H_x$.

To verify the presence of the out-of-plane spin-orbit torque caused by the altermagnetic spin splitting effect in RuO$_2$ and its role in the field-free SOT switching observed in RuO$_2$/Co/Pt/Co/Pt heterostructures, we undertook a detailed investigation of the current-induced effective field. Initially, we examined the deflection variation of the current-induced $R_H$-$H_z$ hysteresis loop under different DC currents. As illustrated in Fig. 3a, in the absence of a transverse magnetic field $H_x$, the anomalous Hall $R_H$-$H_z$ hysteresis loops remain centrosymmetric and show no relative shifts when a positive (+10 mA) and negative (-10 mA) current is applied along the θ = 90° direction (corresponding to the [010] crystal direction) of the RuO$_2$/Co/Pt/Co/Pt sample. Conversely, as depicted in Fig. 3b, applying a stronger positive (+28 mA) and negative (-28 mA) current results in a pronounced shift of the $R_H$ - $H_z$ loop centers to the right and left, respectively. When $H_x$ = 0, no significant shift is detected under small currents. However, when the current increases to ±16 mA, a sudden critical loop shift is observed. Beyond this critical value, the displacement field exhibits an almost linear increase with the current $I$, similar to previously reported systems with an out-of-plane spin torque component[28-32]. Based on the loop shift, an out-of-plane SOT effective field could be extracted from $\Delta H_z^{eff} = H_{shift}(I^+) - H_{shift}(I^-)$ and $H_{shift}(I^\pm) = [|H_C^+(I^\pm)| - |H_C^-(I^\pm)|]/2$. The $\Delta H_z^{eff}$ of + 46 Oe is estimated for ± 28 mA. Similar shifts occur when currents are applied along different θ directions. In contrast, for conventional Co/Pt multilayer structures, there is almost no loop shift at $H_x$ = 0, regardless of the current magnitude. Fig. 3c summarizes the displacement field $\Delta H_z^{eff}$ as a function of current $I$ for RuO$_2$/Co/Pt/Co/Pt and Ta/Co/Pt/Co/Pt when $H_x$ = 0, showing that this current threshold effect occurs at different θ. It can be seen that when the current is applied along the [010] direction (θ=90°), the generated SOT effective field reaches its maximum. The $\Delta H_z^{eff}$ values at other angles exhibit a symmetrical distribution between 0° and 180°, centered around θ = 90°. As the angle deviates further from 90°, the $\Delta H_z^{eff}$ values diminish accordingly. The further the angle deviates from 90°, the smaller the value, mirroring the earlier discussed field-free SOT switching ratio. The observed threshold effect of $\Delta H_z^{eff}$ and the phenomenon of field-free SOT switching serve as compelling evidence for the existence of the z-polarized spin currents in the RuO$_2$/Co/Pt/Co/Pt multilayer system[10,28]. The spin current along the z-direction, arising from the ASSE in RuO$_2$, exhibits distinct anisotropy behavior. Notably, their magnitude is maximized when the current flows along the [010] crystallographic axis direction[8].

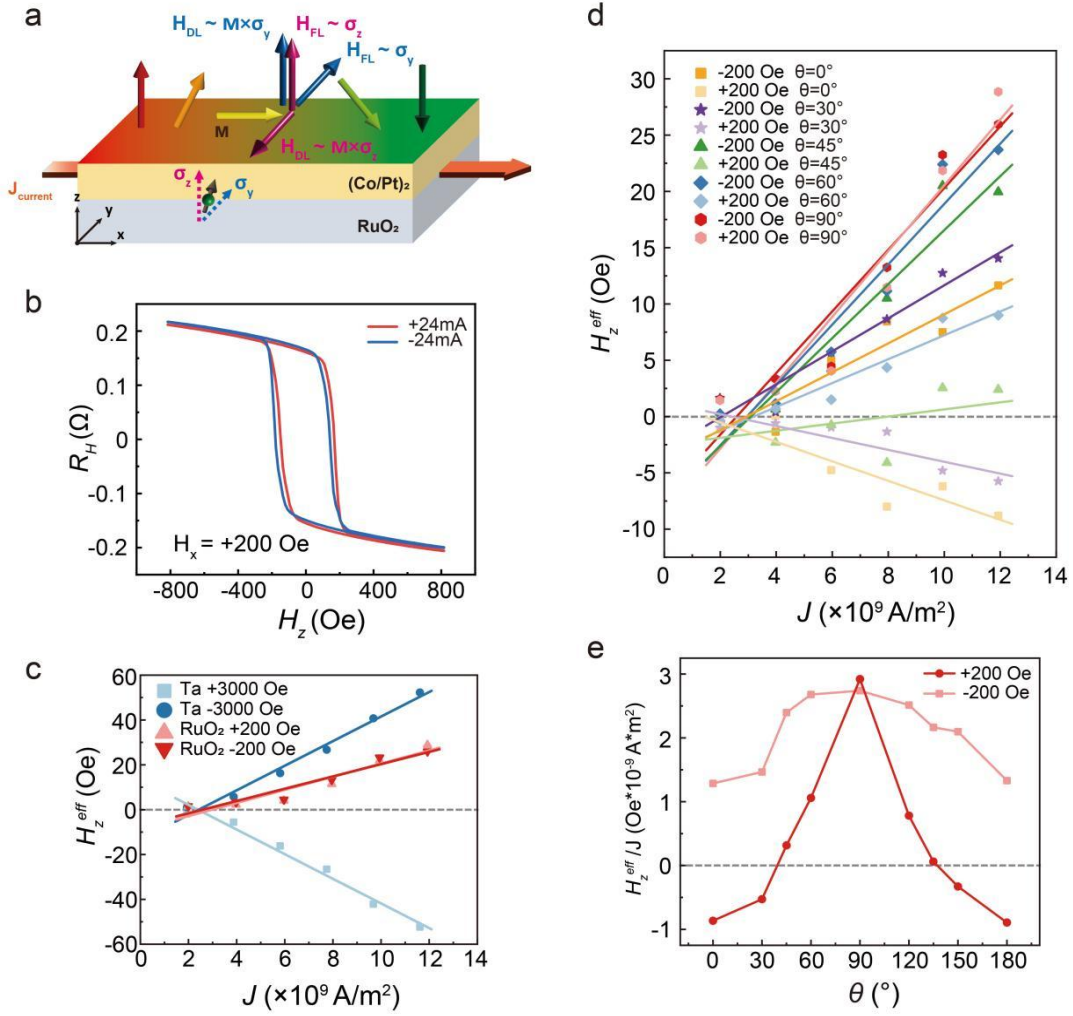

**Fig. 4. Angle-dependent measurements of current-induced effective fields in RuO$_2$ (101). a,** Schematics of magnetization, damping-like (DL) and field-like (FL) fields for RuO$_2$/Co/Pt/Co/Pt structure. **b,** The AHE loops with the applied current of ± 24 mA on RuO$_2$/Co/Pt/Co/Pt at an in-plane field, $H_x$ = + 200 Oe. **c,** A summary of the shift ($\Delta H_z^{eff}$) at different bias currents (*I*) for RuO$_2$/Co/Pt/Co/Pt ($\theta$ = 90°) and Ta/Co/Pt/Co/Pt with $H_x$, respectively. **d,** A summary of the shift ($\Delta H_z^{eff}$) at different bias currents (*I*) for RuO$_2$/Co/Pt/Co/Pt along different current angles $\theta$ with $H_x$ = ± 200 Oe. **e,** A summary of the $\Delta H_z^{eff}/J$ for RuO$_2$/Co/Pt/Co/Pt at different values of $\theta$ with $H_x$ = ± 200 Oe.

In the altermagnetic RuO$_2$, the spin polarization direction is governed by the Néel vector of RuO$_2$, a consequence of the spin splitting effect. The (101)-oriented crystal structure of RuO$_2$, with the Néel vector tilted within the y-z plane, dictates the spin polarization direction along the y- and z-axes. We meticulously investigated the anisotropy of the RuO$_2$ (101)/Co/Pt/Co/Pt by performing current-induced $R_H$-$H_z$ hysteresis loop measurements under different DC currents in the presence of an applied magnetic field. As illustrated in Fig. 4a, our switching measurements consider the multidomain states within the Co layer, where the domain walls exhibit a Néel-type conformation due to the Dzyaloshinskii-Moriya interaction, mediated by spin-orbit coupling with the adjacent Pt layer, thus imparting chirality to the spin texture. In assessing domain wall dynamics, we concentrated on the in-plane magnetization state and the role of spin current. Typically, four effective fields (damping-like (DL) and field-like (FL)) modulate magnetization. The effective perpendicular fields, such as $H_{DL} \sim M \times \sigma_y$ and $H_{FL} \sim \sigma_z$, facilitate domain wall movement, with $H_{FL} \sim \sigma_z$ playing a pivotal role in deterministic magnetization switching. Fig. 4b illustrates the relative movement of the $R_H$-$H_z$ hysteresis loop when positive (+ 24 mA) and negative (- 24 mA) currents are applied along the θ = 90° direction of the RuO$_2$ (101)/Co/Pt/Co/Pt sample (along the [010] crystal direction) for a transverse magnetic field of H$_x$ = + 200 Oe. Fig. 4c presents a comparison of the $\Delta H_z^{eff}$ between RuO$_2$ (101)/Co/Pt/Co/Pt (θ = 90°) and Ta/Co/Pt at $H_x$ = ± 200 Oe and ± 3000 Oe, where the $H_x$ overcome the DMI effective field, respectively. Regardless of the current magnitude, a shift field proportional to the current is observed. The linear relationship between $\Delta H_z^{eff}$ and $I$ provides a quantitative indication of the SOT efficiency, which is $2.74 \times 10^{-9}$ Oe/(A·$m^{-2}$) and $2.92 \times 10^{-9}$ Oe/(A·$m^{-2}$) for RuO$_2$/Co/Pt/Co/Pt and $5.48 \times 10^{-9}$ Oe/(A·$m^{-2}$) and $-5.49 \times 10^{-9}$ Oe/(A·$m^{-2}$) for Ta/Co/Pt/Co/Pt, respectively. Notably, $\Delta H_z^{eff}$ in RuO$_2$(101)/Co/Pt/Co/Pt (θ = 90°) remains isotropic even upon reversal of the applied magnetic field, in contrast to the behavior observed in Ta/Co/Pt/Co/Pt, where $\Delta H_z^{eff}$ reverses with H$_x$, consistent with conventional mechanism resulting from intrinsic SHE[33-35]. Subsequently, we investigated the angular dependence of the $R_H$-$H_z$ hysteresis loops of the RuO$_2$ (101)/Co/Pt/Co/Pt. Fig. 4c illustrates the linear relationship between $\Delta H_z^{eff}$ and current $I$ across various current angles θ, under an applied magnetic field of ± 200 Oe. Interestingly, at θ = 0°, the $\Delta H_z^{eff}$ at unit current density exhibits a symmetric behavior about 0 at H$_x$ = ± 200 Oe. However, this symmetry undergoes a notable transformation as θ incrementally approaches 90°. Specifically, the ratio $\Delta H_z^{eff}/J$ at H$_x$ = + 200 Oe diminishes progressively, transitioning from a negative to a positive value once θ exceeds 45°, and then continues to increase. In contrast, $\Delta H_z^{eff}/J$ at H$_x$ = - 200 Oe remains positive throughout and experiences a steady increase. This intriguing behavior is depicted in Fig. 4d. This behavior arises because the $H_{FL}$ generated by $\sigma_z$ is independent of the direction of the applied magnetic field, being solely contingent upon the magnitude of $\sigma_z$ induced by the current. Further measurements conducted at θ > 90° corroborate this trend, as illustrated in Fig. 4e, elucidating the relationship between $\Delta H_z^{eff}/J$ and θ across a broader angular spectrum.

The observed angular dependence of this phenomenon indicates that the loop shift arises from the synergistic effects of both the conventional Spin Hall Effect (SHE) and the altermagnetic spin splitting effect (ASSE)[6,8,19]. Notably, the ASSE appears to dominate the spin current, particularly when the applied current is oriented along the [010] crystallographic direction (θ = 90°). In this configuration, the spin polarization component $\sigma_z$ exerts a pronounced influence, engendering a substantial field-like effective field, $H_{FL}$, that surpasses the damping-like effective field, $H_{DL}$, associated with $\sigma_y$. This interplay underscores the pivotal role of $\sigma_z$ in driving the enhanced $H_{FL}$, thereby elucidating the underlying mechanics of spin flow modulation in this crystalline context.

In conclusion, we have demonstrated the high quality and excellent structural properties of the RuO$_2$ (101) films grown by magnetron sputtering in this study, alongside their notable electrical conductivity. Our research further reveals that (101)-oriented RuO$_2$ films can effectively generate spin currents with an out-of-plane spin polarization component $\sigma_z$, driven by the altermagnetic spin splitting effect (ASSE). These spin currents exhibit anisotropic behavior, with their magnitude maximized when the charge current flows along the [010] direction. Experimental evidence from SOT switching experiments confirms a strong correlation between spin current direction and crystal orientation in RuO$_2$. Our findings indicate that the z-polarized spin currents induced by ASSE in RuO$_2$ can generate significant out-of-plane anti-damping torque on adjacent magnetic layers, enabling field-free SOT switching. Notably, charge current flowing along the [010] direction achieves the highest field-free switching ratio, maintaining stable SOT switching within an external field range of approximately 400 Oe, demonstrating excellent resistance to magnetic interference under these conditions. The current-induced $R_H$-$H_z$ hysteresis loop measurements underscore the pivotal role of $\sigma_z$ in enhancing spin-torque efficiency, thereby elucidating the underlying mechanisms of spin flow modulation within this crystalline framework. The controllability of spin polarization and the high efficiency of out-of-plane spin current generation position $RuO_2$ as a promising material for next-generation spintronic devices. This research paves the way for practical applications in spin-torque switching technologies, potentially enhancing the functionality, efficiency, and scalability of SOT-MRAM and other spintronic memory and logic devices.

1. Šmejkal, L., Sinova, J. & Jungwirth, T. Emerging Research Landscape of Altermagnetism. *Phys. Rev. X* **12** (2022). https://doi.org/10.1103/PhysRevX.12.040501
2. Šmejkal, L., Sinova, J. & Jungwirth, T. Beyond Conventional Ferromagnetism and Antiferromagnetism: A Phase with Nonrelativistic Spin and Crystal Rotation Symmetry. *Phys. Rev. X* **12** (2022). https://doi.org/10.1103/PhysRevX.12.031042
3. Berlijn, T., Snijders, P. C., Delaire, O. *et al.* Itinerant Antiferromagnetism in RuO2. *Phys. Rev. Lett.* **118**, 077201 (2017). https://doi.org/10.1103/PhysRevLett.118.077201
4. Gonzalez-Hernandez, R., Smejkal, L., Vyborny, K. *et al.* Efficient Electrical Spin Splitter Based on Nonrelativistic Collinear Antiferromagnetism. *Phys. Rev. Lett.* **126**, 127701 (2021). https://doi.org/10.1103/PhysRevLett.126.127701
5. Mattheiss, L. F. Electronic structure of RuO2, OsO2, and IrO2. *Phys. Rev. B* **13**, 2433-2450 (1976). https://doi.org/10.1103/PhysRevB.13.2433
6. Bai, H., Han, L., Feng, X. Y. *et al.* Observation of Spin Splitting Torque in a Collinear Antiferromagnet RuO2. *Phys. Rev. Lett.* **128**, 197202 (2022). https://doi.org/10.1103/PhysRevLett.128.197202
7. Bose, A., Schreiber, N. J., Jain, R. *et al.* Tilted spin current generated by the collinear antiferromagnet ruthenium dioxide. *Nat. Electron.* **5**, 267-274 (2022). https://doi.org/10.1038/s41928-022-00744-8
8. Guo, Y., Zhang, J., Zhu, Z. *et al.* Direct and Inverse Spin Splitting Effects in Altermagnetic RuO2. *Adv. Sci.*, e2400967 (2024). https://doi.org/10.1002/advs.202400967
9. MacNeill, D., Stiehl, G. M., Guimaraes, M. H. D. *et al.* Control of spin–orbit torques through crystal symmetry in WTe2/ferromagnet bilayers. *Nat. Phys.* **13**, 300-305 (2016). https://doi.org/10.1038/nphys3933
10. Baek, S. C., Amin, V. P., Oh, Y. W. *et al.* Spin currents and spin-orbit torques in ferromagnetic trilayers. *Nat. Mater.* **17**, 509-513 (2018). https://doi.org/10.1038/s41563-018-0041-5
11. Xie, H., Chen, X., Zhang, Q. *et al.* Magnetization switching in polycrystalline Mn(3)Sn thin film induced by self-generated spin-polarized current. *Nat. Commun.* **13**, 5744 (2022). https://doi.org/10.1038/s41467-022-33345-2
12. Song, C., Zhang, R., Liao, L. *et al.* Spin-orbit torques: Materials, mechanisms, performances, and potential applications. *Prog. Mater. Sci.* **118** (2021). https://doi.org/10.1016/j.pmatsci.2020.100761
13. Han, X., Wang, X., Wan, C. *et al.* Spin-orbit torques: Materials, physics, and devices. *Appl. Phys. Lett.* **118** (2021). https://doi.org/10.1063/5.0039147
14. Du, A., Zhu, D., Cao, K. *et al.* Electrical manipulation and detection of antiferromagnetism in magnetic tunnel junctions. *Nat. Electron.* **6**, 425-433 (2023). https://doi.org/10.1038/s41928-023-00975-3
15. Wang, F., Shi, G., Kim, K. W. *et al.* Field-free switching of perpendicular magnetization by two-dimensional PtTe(2)/WTe(2) van der Waals heterostructures with high spin Hall conductivity. *Nat. Mater.* (2024). https://doi.org/10.1038/s41563-023-01774-z
16. Baltz, V., Manchon, A., Tsoi, M. *et al.* Antiferromagnetic spintronics. *Rev. Mod. Phys.* **90** (2018). https://doi.org/10.1103/RevModPhys.90.015005
17. Jungwirth, T., Marti, X., Wadley, P. & Wunderlich, J. Antiferromagnetic spintronics. *Nat. Nanotechnol.* **11**, 231-241 (2016). https://doi.org/10.1038/nnano.2016.18

18   Shao, Q., Li, P., Liu, L. *et al.* Roadmap of Spin–Orbit Torques. *IEEE T. Magn.* **57**, 1-39 (2021). https://doi.org/10.1109/tmag.2021.3078583

19   Bai, H., Zhang, Y. C., Zhou, Y. J. *et al.* Efficient Spin-to-Charge Conversion via Altermagnetic Spin Splitting Effect in Antiferromagnet RuO2. *Phys. Rev. Lett.* **130**, 216701 (2023). https://doi.org/10.1103/PhysRevLett.130.216701

20   Zhang, Z., Lu, X., Yan, Y. *et al.* Direct observation of spin polarization in epitaxial Fe3O4(001)/MgO thin films grown by magnetron sputtering. *Appl. Phys. Lett.* **120** (2022). https://doi.org/10.1063/5.0091241

21   Nunn, W., Nair, S., Yun, H. *et al.* Solid-source metal–organic molecular beam epitaxy of epitaxial RuO2. *APL Mater.* **9** (2021). https://doi.org/10.1063/5.0062726

22   Meng, L. Raman spectroscopy analysis of magnetron sputtered RuO2 thin films. *Thin Solid Films* **442**, 93-97 (2003). https://doi.org/10.1016/s0040-6090(03)00953-2

23   Mar, S. Y., Chen, C. S., Huang, Y. S. & Tiong, K. K. Characterization of RuO2 thin films by Raman spectroscopy. *Appl. Surf. Sci.* **90**, 497-504 (1995). https://doi.org/10.1016/0169-4332(95)00177-8

24   Ryden, W. D., Lawson, A. W. & Sartain, C. C. Electrical Transport Properties of IrO2 and RuO2. *Phys. Rev. B* **1**, 1494-1500 (1970). https://doi.org/10.1103/PhysRevB.1.1494

25   Liu, L., Pai, C. F., Li, Y. *et al.* Spin-torque switching with the giant spin Hall effect of tantalum. *Science* **336**, 555-558 (2012). https://doi.org/10.1126/science.1218197

26   Liu, L., Lee, O. J., Gudmundsen, T. J. *et al.* Current-induced switching of perpendicularly magnetized magnetic layers using spin torque from the spin Hall effect. *Phys. Rev. Lett.* **109**, 096602 (2012). https://doi.org/10.1103/PhysRevLett.109.096602

27   Kong, W. J., Wan, C. H., Wang, X. *et al.* Spin-orbit torque switching in a T-type magnetic configuration with current orthogonal to easy axes. *Nat. Commun.* **10**, 233 (2019). https://doi.org/10.1038/s41467-018-08181-y

28   Liu, L., Zhou, C., Shu, X. *et al.* Symmetry-dependent field-free switching of perpendicular magnetization. *Nat. Nanotechnol.* **16**, 277-282 (2021). https://doi.org/10.1038/s41565-020-00826-8

29   Wang, M., Zhou, J., Xu, X. *et al.* Field-free spin-orbit torque switching via out-of-plane spin-polarization induced by an antiferromagnetic insulator/heavy metal interface. *Nat. Commun.* **14**, 2871 (2023). https://doi.org/10.1038/s41467-023-38550-1

30   Jin, T., Lim, G. J., Poh, H. Y. *et al.* Spin Reflection-Induced Field-Free Magnetization Switching in Perpendicularly Magnetized MgO/Pt/Co Heterostructures. *ACS Appl. Mater. Interfaces* **14**, 9781-9787 (2022). https://doi.org/10.1021/acsami.1c22061

31   Hu, S., Shao, D. F., Yang, H. *et al.* Efficient perpendicular magnetization switching by a magnetic spin Hall effect in a noncollinear antiferromagnet. *Nat. Commun.* **13**, 4447 (2022). https://doi.org/10.1038/s41467-022-32179-2

32   Lee, K. J., Liu, Y., Deac, A. *et al.* Spin transfer effect in spin-valve pillars for current-perpendicular-to-plane magnetoresistive heads (invited). *J. Appl. Phys.* **95**, 7423-7428 (2004). https://doi.org/10.1063/1.1682872

33   Karube, S., Tanaka, T., Sugawara, D. *et al.* Observation of Spin-Splitter Torque in Collinear Antiferromagnetic RuO2. *Phys. Rev. Lett.* **129**, 137201 (2022). https://doi.org/10.1103/PhysRevLett.129.137201

34   Li, Z., Lu, X., Zhang, Z. *et al.* Efficient spin–orbit torque switching in perpendicularly magnetized CoFeB facilitated by Fe2O3 underlayer. *Appl. Phy. Lett.* **123** (2023). https://doi.org/10.1063/5.0163034

35   Pai, C.-F., Mann, M., Tan, A. J. & Beach, G. S. D. Determination of spin torque efficiencies in heterostructures with perpendicular magnetic anisotropy. *Phys. Rev. B* **93** (2016). https://doi.org/10.1103/PhysRevB.93.144409